\documentclass[structabstract]{aa}  
\usepackage{graphicx}
\usepackage{natbib}
\usepackage{amsmath}
\usepackage{rotating}
\usepackage{lscape}
\usepackage{color}
\usepackage{latexsym}
\usepackage{amsfonts}
\usepackage{multirow}
\usepackage{rotating}
\usepackage{graphicx}
\usepackage{epsfig}
\usepackage{amssymb,amsbsy}
\bibpunct{(}{)}{;}{a}{}{,}
\newcommand{\ie}{$i.e.,\;$}
\newcommand{\eg}{$e.g.,\;$}

\usepackage{txfonts}
\begin{document}
    \title{Measuring the level of nuclear activity in Seyfert galaxies and the unification scheme}

   \author{Veeresh Singh\inst{1,2}\fnmsep\thanks{veeresh@iiap.res.in}, Prajval Shastri\inst{1}
          \and Guido Risaliti\inst{3,4} }
   \institute{Indian Institute of Astrophysics, Bangalore 560034, India
           \and
              Department of Physics, University of Calicut, Calicut 673635, India 
            \and
              INAF-Osservatorio di Arcetri, Largo E. Fermi 5, I-50125 Firenze, Italy 
         \and
             Harvard-Smithsonian Center for Astrophysics, 60 Garden St. Cambridge, MA 02138, USA
 }

   \date{Received xxxx xx, xxxx; accepted xxxx xx, xxxx}

 
  \abstract
{The unification scheme of Seyfert galaxies hypothesizes that Seyfert type 1s and type 2s are intrinsically similar and the observed differences 
between the two subtypes are solely due to the differing orientations of toroidal-shaped obscuring material around the AGN. 
In the framework of the unification scheme, both the Seyfert subtypes are expected to show similar intrinsic nuclear properties, such as 
the absorption-corrected AGN X-ray luminosity, bolometric luminosity, accretion rate and the mass of the supermassive black hole.}
{To test the predictions of the Seyfert unification scheme, we make statistical comparison of the distributions of: 
(i) the absorption-corrected 2.0 - 10 keV X-ray luminosities, (ii) the bolometric luminosities, (iii) the black hole masses and, (iv) the Eddington ratios, of Seyfert type 1s and type 2s.}
{We use an optically selected Seyfert sample in which type 1s and type 2s are matched in properties that are independent to the orientation of 
the obscuring torus, the AGN axis and the host galaxy.}
{The distributions of the absorption-corrected 2.0 - 10 keV X-ray luminosities (L$_{\rm 2.0 - 10~keV}^{\rm c}$), 
the bolometric luminosities (L$_{\rm Bol}$), the black masses (M$_{\rm BH}$) and, the Eddington ratios ($\lambda$) are statistically similar for the two Seyfert subtypes, 
consistent with the orientation and obscuration based Seyfert unification scheme. 
The Eddington ratio distributions suggest that both the Seyfert subtypes are accreting at sub-Eddington level with wide span of Eddington ratios 
{\ie}10$^{-4}$ - 10$^{-1}$.}
    {}


\authorrunning{Singh et al.}
   \maketitle              
%

\section{Introduction}
Seyfert galaxies are categorized as nearby, low-luminosity, radio-quiet AGNs powered by accretion onto supermassive black holes.
Seyfert galaxies are classified mainly into two subclasses named as type~1 and type~2, based on the presence and absence 
of broad emission lines in their optical spectra, respectively \citep{Ant93}. 
Seyfert unification scheme hypothesizes that the two subtypes are intrinsically similar {\ie}belong to the same parent population 
and appear different solely due to the differing orientations of the obscuring material having toroidal geometry around the AGN. 
When the plane of the obscuring torus is along the observer's line-of-sight {\ie}edge-on view (type~2s), 
the central engine and broad line region clouds are hidden, while in pole-on view (type~1s) the central engine 
as well as broad line region clouds are directly seen \citep{Ant85,Ant93,Urry95}.
\par
There have been several attempts to investigate the validity of Seyfert unification scheme, giving both consistent as well as inconsistent results. 
Results such as the presence of broad emission lines in the polarized optical and infrared spectra of many 
Seyfert 2s \citep{Moran2000}, the biconical structure of the narrow line region \citep{Wilson96}, 
systematically higher X-ray absorbing column density in Seyfert type 2s \citep{Cappi06} and
similar nuclear radio properties of both the subtypes \citep{Lal11} are consistent with the unification scheme. 
While, results such as the absence of hidden Seyfert~1 nuclei in several Seyfert~2s \citep{Tran01,Tran03}, 
Seyfert~1s being preferentially hosted in galaxies of earlier Hubble type \citep{Malkan98}, 
lack of X-ray absorption in some Seyfert~2s \citep{Panessa02}, and Seyfert~2s having a higher propensity of nuclear starbursts 
\citep{Buchanan06} are inconsistent with the unification scheme. 
Sample selection is arguably the most important issue in testing the predictions of Seyfert unification scheme and 
several results inconsistent to the scheme can be explained as due to the biases inherent in the samples \citep{Antonucci02}. 
Indeed, it has been shown that the Seyfert samples selected from optical and UV surveys tend to be biased 
against obscured and faint sources \citep{Ho01}. 
The obscuring torus absorbs the optical, UV photons emanating from AGN and reradiates in IR. 
Also, emission at IR wavelengths does not suffer large extinction, therefore, IR selected samples are likely to be less susceptible to selection biases \citep{Spinoglio89}. 
However, IR selected samples can be biased towards unusually dusty and sources with higher level of nuclear star formation \citep{Buchanan06}.
Seyfert samples selected from X-ray surveys have also been used to test the predictions of the Seyfert unification 
\citep{Awaki91,Smith96,Turner97,Bassani99}.
However, studies based on [OIII] selected samples have shown that the X-ray (E $<$ 10 keV) selected samples are biased 
against heavily obscured AGNs \citep{Risaliti99,Heckman05}.
Hard X-ray photons can transmit through the obscuring material and 
therefore, samples based on hard X-ray surveys are expected to be least biased. 
However, Seyfert samples based on {\em INTEGRAL} and {\em Swift}/BAT surveys preferentially contain relatively large number 
of high luminosity and less absorbed Seyferts \citep{Tueller08,Treister09,Beckmann09}, possibly due to less effective area which limits the sensitivity 
only to bright sources ($\sim$10$^{-11}$ erg s$^{-1}$ cm$^{-2}$). Therefore, samples based on hard X-ray surveys are likely to biased against 
heavily obscured Compton-thick and low luminosity AGNs \citep{Heckman05,Wang09}.
Thus, in general, samples derived from flux limited surveys tend to bias against obscured and faint sources.
\\
The quest of testing the validity and limitations of Seyfert unification scheme with more improved and 
well defined samples continues ({\eg}\cite{Cappi06,Dadina08,Horst08,Ricci11,Brightman11a}), however, 
issues related to the sample selection still remain. 
It has been argued that the samples based on the properties that are independent to 
the orientation of the obscuring torus are more appropriate to test the unification scheme \citep{Schmitt03b,Lal11,Singh11}. 
Such samples are less prone to the biases that are generally inherent in the samples derived from flux limited surveys.
\cite{Lal11} have presented a Seyfert sample based on orientation-independent properties and 
the two Seyfert subtypes of this sample are matched in the orientation-independent properties.
Keeping above sample selection issues in mind, we use \cite{Lal11} sample with 
the aim to test the Seyfert unification scheme by making statistical comparison of the distributions of 
nuclear properties such as the absorption-corrected 2.0 - 10 keV X-ray luminosities (L$_{\rm 2.0~-~10~keV}^{\rm c}$), 
the bolometric luminosities (L$_{\rm Bol}$), the black hole masses (M$_{\rm BH}$) and the Eddington ratios ($\lambda$) 
of the Seyfert type 1s and type 2s. 
\\
The structure of this paper is as following. 
In section 2, we discuss the sample details. In section 3, we discuss the empirical relations that we used to derive the bolometric 
luminosities of the sample sources. A brief note on Eddington ratios is given in section 4. In section 5, we discuss the comparisons of 
the absorption-corrected luminosities, the bolometric luminosities, the black hole masses and the Eddington ratios of the two Seyfert subtypes.
Throughout the paper we have assumed cosmological parameters H$_{\rm 0}$ = 71 km$^{-1}$ Mpc$^{-1}$, ${\Lambda}_{\rm m}$ = 0.27, and ${\Lambda}_{\rm vac}$ = 0.73.
\section{The sample}
We use the Seyfert sample of \cite{Lal11} that is consist of 20 (10 type 1s and 10 type 2s) Seyfert galaxies. 
In this sample, Seyfert galaxies are defined as radio-quiet ($\frac{\rm F_{5~GHz}}{\rm F_{B-band}}$ $<$ 10; \cite{Kellerman89}), 
low optical luminosity AGN (M$_{\rm B}$ $>$ -23; \cite{Schmidt83}), hosted 
in spiral or lenticular galaxies \citep{Weedman77}. 
The sample selection is based on the properties that do not depend on the orientation of the obscuring torus, the AGN axis and the host galaxy. 
The orientation-independent properties used for the sample selection are: 
cosmological redshift, [OIII] $\lambda$5007$\mbox {\AA}$ line 
luminosity, Hubble stage of the host galaxy, absolute stellar magnitude of the host galaxy and absolute bulge magnitude.
All these properties are also intimately linked to the evolution of AGN as well as host galaxy. 
The sample is selected such that the two Seyfert subtypes have matched distributions in the orientation-independent properties. 
We refer readers to see \cite{Lal11} for greater details on the sample selection. 
The sample was formulated to study the nuclear radio properties of Seyferts and is constrained by 
the Very Large Baseline Interferometer (VLBI) observing feasibility. 
\cite{Lal11} noted that 54 Seyferts met VLBI observing feasibility criterion and 29 of these 54 had all the required 
orientation-independent parameters in the literature. From these 29, they picked 20 Seyferts such that the two Seyfert subtypes had matched 
distributions in the orientation-independent parameters. The sources which were deviating in matching the distributions of the 
orientation-independent parameters were left out. In other words, the two Seyfert subtypes should lie within the same range of values for 
a given parameter to enter into the sample. Also, it was ensured that in a given bin of a parameter distribution, type 1s do not outnumber 
the type 2s and vice-versa. Indeed, the selected 20 Seyferts are not unique or complete in any aspect. 
Nevertheless, the sample is sufficiently qualified to test the predictions of Seyfert unification scheme 
at any wavelength regime since it is based on the orientation-independent properties, and ensure that the two Seyfert subtypes 
are intrinsically similar within the framework of the unification scheme.  
Furthermore, it is possible to enlarge the sample size by relaxing the VLBI observing feasibility criterion, but keeping 
the criterion of similar distributions for the two Seyfert subtypes in the orientation-independent parameters, intact. 
However, we emphasize that the more important issue is the sample selection and not the sample size. 
\\
The criterion of similar distributions of the orientation-independent parameters for the two subtypes and VLBI observing feasibility 
may introduce a positive bias towards the sources having similar AGN and host galaxies properties. 
Therefore, this sample may not be claimed as a complete representative of the entire class of Seyfert galaxies. 
However, more importantly, matched distributions are essential to ensure that we are not comparing entirely intrinsically different 
sources selected from different parts of the evolution function (luminosity, bulge mass, Hubble type, redshift). 
Since the AGN and host galaxy properties are likely to affect the AGN surrounding environment {\eg}opening angle of the obscuring torus 
decreases with the increase in AGN luminosity \citep{Ueda03,Hasinger08}, formation of BLR clouds may depend on the AGN luminosity 
\citep{Laor03,Nicastro2000}, in low luminosity AGNs accretion rate increases from early-type to late-type galaxies \citep{Ho09}. 
Therefore, the samples that have large variations in the AGN and host galaxy properties are 
expected to show the effect of these variations in the observed properties other than the differences 
owing to the differing orientations of the obscuring torus. \cite{Lal11} Seyfert sample based on the matched distributions of the 
orientation-independent parameters, minimizes the impact of differences caused by the differing properties of 
the AGN as well as host galaxy.
\\
Using the same sample we have shown in our previous paper (\cite{Singh11}, hereafter paper-I) that the comparisons of X-ray properties 
{\eg}distributions of the observed X-ray luminosities, the absorbing column densities, the equivalent widths of Fe K$\alpha$ line and the flux ratios 
of 2.0 - 10 keV hard X-ray to [OIII] line emission of the two Seyfert subtypes, are consistent with the Seyfert unification scheme.
\section{Estimating bolometric luminosity}
In order to calculate the bolometric luminosity of a source one requires complete spectral energy distribution (SED) of the source
over the entire electromagnetic spectrum. However, in practice complete SED is not available for most of the AGNs.
In the literature, both nuclear 2.0 - 10 keV X-ray luminosity as well as [OIII] $\lambda$5007${\mbox {\AA}}$ line luminosity have been used 
to estimate the bolometric luminosity \citep{Heckman05,Kauffmann09,Lamastra09}. 
In following subsections we discuss the empirical relations that we use to estimate the bolometric luminosities for our sample sources.
\subsection{Using X-ray luminosity} 
Different studies in the literature have suggested different correction factors to estimate the bolometric luminosity from 
the 2.0 - 10 keV X-ray luminosity \citep{Elvis94,Ho08}.
However, SED of an AGN vary with accretion rate and 
therefore it may not be appropriate to use a single correction factor to 
estimate the bolometric luminosity (L$_{\rm Bol}$) from the nuclear X-ray luminosity (L$_{\rm 2.0~-~10~keV}$).
We estimated the bolometric luminosity from absorption-corrected 2.0 - 10 keV X-ray luminosity using the empirical relation given in \cite{Marconi04}.
\begin{equation} 
\log \left(\frac{\rm L_{\rm Bol}}{\rm L_{\rm 2.0~-~10~keV}^{\rm c}}\right)=1.54+0.24\mathcal {L}+0.012\mathcal {L}^2-0.0015\mathcal {L}^3
\end{equation}
where $\mathcal {L}=\log {\rm L_{\rm Bol}}-12$, and L$_{\rm Bol}$, L$_{\rm 2.0 - 10~keV}^{\rm c}$ are in units of L$_{\odot}$.
\cite{Marconi04} derived this bolometric correction using an average intrinsic spectral energy distribution template for radio-quiet 
AGNs and also accounted for the variations in AGN SEDs by using the well-known anti-correlation between the optical-to-X-ray spectral index 
(${\alpha}_{\rm OX}$) and 2500 ${\mbox \AA}$ luminosity.
We obtain the absorption-corrected 2.0~-~10~keV X-ray luminosities (L$_{\rm 2.0 - 10~keV}^{\rm c}$) of our sample sources using the 0.5 - 10 keV 
{\em XMM-N} X-ray spectral fits (see, Paper-I).
For Compton-thin sources, the absorbing column densities are accurately known and thus the absorption-corrected fluxes/luminosities 
can be obtained by fixing the equivalent hydrogen column density (N$_{\rm H}$) equal to `0' in the best X-ray spectral fitting models. 
However, the correction for X-ray absorption in case of the Compton-thick sources is non-trivial since 
photoelectric absorption cut-off and hence absorbing column density is not measurable by the X-ray observations that are limited up to 10 keV, 
{\eg}{\em XMM-Newton}, {\em Chandra} and {\em ASCA} observations.
In order to get the absorption-corrected X-ray fluxes for Compton-thick sources we have used flux diagnostic ratio which is based 
on measuring the X-ray flux against [OIII] $\lambda$5007$\mbox {\AA}$ flux, {\ie}Rx = $\frac{\rm F_{2.0-10.0~keV}}{\rm F_{[OIII]}}$ \citep{Bassani99,Panessa06}.
We estimate the absorption-corrected 2.0 - 10.0 keV X-ray flux/luminosity for a Compton-thick source 
by multiplying the observed flux/luminosity with a correction factor f$_{\rm cor}$ that is approximated as the ratio of the medians
of Rx$_{\rm Compton-thin}$ to Rx$_{\rm Compton-thick}$ 
({\ie}f$_{\rm cor}$ = $\frac{\rm Rx_{\rm Compton-thin,~median}}{\rm Rx_{Compton-thick,~median}}$, \cite{Gonz09}), where 
Rx$_{\rm Compton-thin}$ is $\frac{\rm F_{(2.0-10.0~keV)cor}}{\rm F_{[OIII]cor}}$ ({\ie}the ratio of the absorption-corrected 2.0 - 10.0 keV flux to the reddening corrected [OIII] $\lambda$5007$\mbox {\AA}$ flux for a  Compton-thin source), while Rx$_{\rm Compton-thick}$ is $\frac{\rm F_{(2.0-10.0~keV)obs}}{\rm F_{[OIII]cor}}$ ({\ie}the ratio of the observed 2.0 - 10.0 keV flux to the reddening corrected [OIII] $\lambda$5007$\mbox {\AA}$ flux for a Compton-thick sources). 
In paper-I we report that in our sample of 20 Seyferts, 7/10 type 2 Seyferts are Compton-thick and 
13 Seyferts (10 type 1s and 3 type 2s) are Compton-thin. 
Using the median values of Rx$_{\rm Compton-thin}$ and Rx$_{\rm Compton-thick}$ in this sample, we 
estimated the absorption correction factor f$_{\rm cor}$ $\sim$ 70, which is close to the one reported in \cite{Panessa06} by using 
similar flux diagnostic method but considering the flux ratios of Seyfert type 1s and type 2s rather than that of Compton-thick and Compton-thin sources.
We caution that the true value of the correction factor (f$_{\rm cor}$) may be different for different sources depending upon 
the true value of the absorbing column density. Moreover, we note that our approximated value for the correction factor F$_{\rm cor}$ is 
not too different than that for the typical Compton-thick Seyferts \citep{Levenson06,Brightman11a} and therefore, 
it is unlikely to affect our analysis statistically. 

\subsection{Using [OIII] $\lambda$5007${\mbox {\AA}}$ line luminosity}
The [OIII] $\lambda$5007${\mbox {\AA}}$ line emission originates in the narrow line region and its luminosity correlates with 
the total AGN power \citep{Mulchaey94,Heckman05}. 
Also unlike X-ray emission which can be heavily absorbed ({\eg}in Compton-thick AGNs), [OIII] line emission is not much 
affected by dust obscuration from the torus since it originates in the narrow line region which lies outside the torus.
\cite{Heckman04} reported that for type 1 Seyfert nuclei the average bolometric correction to the 
extinction uncorrected [OIII] luminosity is (L$_{\rm Bol}$/L$_{\rm [OIII]}$) $\sim$ 3500, with a variance of 0.38.
However, extinction for the [OIII] emission caused by dust present in NLR itself can be substantial and therefore 
extinction-corrected [OIII] luminosity (L$^{\rm c}_{\rm OIII}$) is a more direct indicator of the nuclear luminosity.
To estimate the bolometric luminosity from the [OIII] line luminosity we have used the correction factors given in \cite{Lamastra09}
that are estimated by following a method similar to \cite{Heckman04}, but using the extinction-corrected [OIII] 
luminosity instead of the observed one. 
\cite{Lamastra09} reported the luminosity dependent bolometric correction factors (C$_{\rm [OIII]}$) 87, 142, 454 for the 
[OIII] luminosity ranges logL$_{\rm [OIII]}$ = 38 - 40, 40 - 42, 42 - 44, respectively. 
The luminosity-dependent [OIII] correction factors (C$_{\rm [OIII]}$) are obtained by using L$_{\rm X}$ - L$_{\rm [OIII]}^{\rm c}$ correlation fit 
and the luminosity-dependent X-ray bolometric correction of \cite{Marconi04}.
For the sources which do not have extinction-corrected [OIII] luminosities we have used \cite{Heckman04} relation to derive 
the bolometric luminosities.
\section{Eddington ratio}
Eddington luminosity (L$_{\rm Edd}$) $\simeq$ 1.3 $\times$ 10$^{38}$ (M$_{\rm BH}$/M$_{\sun}$) is an upper limit of the 
luminosity produced by a black hole of mass M$_{\rm BH}$. The Eddington limit is a physical limit at which the outward radiation pressure 
from the accreting matter balances the inward gravitational pressure exerted by the black hole.
The mass accretion rate can be parametrized by Eddington ratio ($\lambda$ = L$_{\rm Bol}$/L$_{\rm Edd}$) that is the ratio of 
the bolometric luminosity (L$_{\rm Bol}$) to the Eddington luminosity (L$_{\rm Edd}$) for a given black hole mass (M$_{\rm BH}$). 
The Eddington ratio can be expressed as 
\begin{equation}
\lambda = \frac{\rm L_{\rm bol}}{\rm L_{\rm Edd}} \simeq 0.1 \left(\frac{\rm L_{\rm bol}}{1.4 \times 10^{44}~{\rm erg~s^{-1}}}\right) \left(\frac{\rm M_{\rm BH}}{\rm 10^{7} M_{\odot}}\right)
\end{equation}
\citep{Wang07}.
For all of our sample sources (except for MRK 1218), 
the estimated black hole mass is available in the literature. The black hole masses are estimated using different methods such as 
reverberation mapping, M-$\sigma$ relation \citep{Gebhardt2000,Ferrarese2000}. 
In order to calculate Eddington ratio, we need bolometric luminosity (L$_{\rm Bol}$) that 
we estimate using the absorption-corrected 2.0 - 10 keV X-ray luminosity as well as 
the [OIII] line luminosity. Thus, for a source, we have two values of the Eddington ratio, one is ${\lambda}_{\rm X}$, where L$_{\rm Bol}^{\rm X}$ is used for the 
bolometric luminosity and the other one is ${\lambda}_{\rm [OIII]}$, where L$_{\rm Bol}^{\rm [OIII]}$ is used for the bolometric luminosity. 
Figure 3 shows that the bolometric luminosities obtained from the X-ray (L$_{\rm Bol}^{\rm X}$) and 
the [OIII] luminosities (L$_{\rm Bol}^{\rm [OIII]}$) are similar.
There are three sources ({\ie}MRK 231, MRK 1 and NGC 5135) which show more than one order of magnitude difference between 
L$_{\rm Bol}^{\rm X}$ and L$_{\rm Bol}^{\rm [OIII]}$. One possible reason for it may be that L$_{\rm Bol}^{\rm [OIII]}$ values for 
these sources are overestimated due to large scatter involved in L$^{\rm c}_{\rm [OIII]}$ $-$ L$_{\rm Bol}$ relation 
{\ie}in bolometric correction factor C[OIII] \citep{Lamastra09}. Starburst contamination to L$_{\rm [OIII]}$ may also attribute to the overestimation of 
L$_{\rm Bol}^{\rm [OIII]}$ as some of these sources {\eg}MKR 231 \citep{Taylor99}, NGC 5135 \citep{Gonzalez98} are known to host 
circumnuclear starburst. Furthermore, L$_{\rm [OIII]}$ is only an indirect estimator of the nuclear luminosity and 
it depends on various factors such as on the geometry of the system, on the amount of gas, and on any possible shielding effect that may affect 
the ionizing radiation seen by the NLR.
\begin{table*}
\centering
\begin{minipage}{140mm}
\caption{Absorption-corrected 2.0 - 10 keV X-ray luminosities, bolometric luminosities, black hole masses and Eddington ratios for two Seyfert subtypes.}
\begin{tabular} {cccccccccc} 
\hline
Name  & Seyfert       &logL$_{\rm 2.0 - 10~keV}^{\rm c}$ & logL$_{\rm [OIII]}^{\rm c}$ & logL$_{\rm Bol}^{\rm X}$ & logL$_{\rm Bol}^{\rm [OIII]}$ & log M$_{\rm BH}$ & Ref & log${\lambda}_{\rm X}$ & log${\lambda}_{\rm [OIII]}$ \\
      &   type      & (erg s$^{-1}$) & (erg s$^{-1}$) &  (erg s$^{-1}$) &  (erg s$^{-1}$) & (M$_{\odot}$) &  &  &   \\  \hline
MCG+8-11-11 &  1 & 43.79  & 42.43  &  45.25   & 45.09    &  8.06 & 6  & -1.92   &  -2.09    \\
MRK 1218    &  1 & 42.69  & 41.61  &  43.87   & 43.76    &  .... &  ....  &  ....   &   ....   \\
NGC 2639    &  1 & 41.92  & 40.86  &  42.94   & 43.02    &  8.02 & 3  &  -4.20   &  -4.12    \\
NGC 4151    &  1 & 42.40  & 41.39  &  43.51   & 43.54    &  7.18 & 2  &  -2.79   &  -2.75    \\
MRK 766     &  1 & 42.50  & 41.84  &  43.62   & 43.99    &  6.64 & 4  &  -2.13   &  -1.76    \\
MRK 231     &  1 & 42.53  & 41.96$^{\dagger}$ &  43.66   & 45.51    &  7.24 & 5  &  -2.69   &  -0.85    \\
ARK 564     &  1 & 43.37  & 41.38$^{\dagger}$ &  44.71   & 44.93    &  6.50 & 4  &  -0.91   &  -0.69    \\
NGC 7469    &  1 & 43.45  & 42.96  &  44.82   & 45.62    &  6.84 & 7  &  -1.14   &  -0.34    \\
MRK 926     &  1 & 44.20  & 42.25$^{\dagger}$ &  45.79   & 45.79    &  7.14 & 6  &  -0.47   &  -0.46    \\
MRK 530     &  1 & 43.70  & 40.98$^{\dagger}$ &  45.14   & 44.53    &  8.08 & 7  &  -2.06   &  -2.67    \\
MRK 348     &  2 & 43.45  & 42.19  &  44.81   & 44.85    &  7.21 & 7  &  -1.51   &  -1.47    \\
MRK 1       &  2 & 42.50  & 42.39  &  43.63   & 45.04    &  7.16 & 7  &  -2.64   &  -1.23    \\
NGC 2273    &  2 & 42.74  & 41.16  &  43.92   & 43.82    &  7.30 & 7  &  -2.50   &  -2.60    \\
MRK 78      &  2 & 44.09  & 43.29  &  45.64   & 45.94    &  7.87 & 7  &  -1.34   &  -1.04    \\
NGC 5135    &  2 & 42.68  & 42.42  &  43.85   & 45.08    &  7.35 & 1  &  -2.62   &  -1.39    \\
MRK 477     &  2 & 44.44  & 43.44  &  46.11   & 46.10    &  7.20 & 1  &  -0.21   &  -0.22    \\
NGC 5929    &  2 & 42.04  & 40.98  &  43.07   & 43.64    &  7.25 & 7  &  -3.29   &  -2.72    \\
NGC 7212    &  2 & 43.90  & 42.80  &  45.39   & 45.46    &  7.47 & 1  &  -1.19   &  -1.12    \\
MRK 533     &  2 & 43.91  & 42.64  &  45.41   & 45.30    &  7.56 & 7  &  -1.27   &  -1.38    \\
NGC 7682    &  2 & 43.07  & 41.76  &  44.33   & 44.42    &  7.28 & 7  &  -2.06   &  -1.98    \\ \hline
\end{tabular}
\\
Notes: Col. (1), source name; col. (2), Seyfert subtype; col. (3), absorption-corrected 2.0 - 10 keV X-ray luminosity; 
col. (4), extinction-corrected [OIII] luminosity; col. (5), bolometric luminosity derived from 2.0 - 10 keV X-ray luminosity; 
col. (6), bolometric luminosity derived from [OIII] line luminosity; col. (7), black hole masses; col. (8), references for black hole masses; 
col. (9) Eddington ratio where the estimated bolometric luminosity is derived from X-ray luminosity; 
col (10), Eddington ratio where the estimated bolometric luminosity is derived from [OIII] line luminosity. \\
$\dagger$ : [OIII] luminosity is uncorrected for extinction as the Balmer decrement value is unavailable. \\
References- 1: \cite{Bian07}, 2: \cite{Kaspi2000a}, 3: McElory (1995), 4: \cite{Merloni03}, 5: \cite{Dasyra06}, 6: \cite{Wang07}, 7: \cite{Woo02}.
\end{minipage}
\end{table*}
\begin{table} 
\caption{Medians and Kolmogorov-Smirnov two sample tests for the statistical comparison of various distributions of the two Seyfert subtypes}
\hspace {1.0cm} 
\begin{tabular} {ccccccc}
\hline
Distribution                 &    \multicolumn{2}{c}{Median}      & D       &  p value   \\
                             & Type 1s & Type 2s      &                &                    \\ \hline 
 log L$_{\rm 0.5 - 2.0~keV}^{\rm c}$  &  43.15  & 43.30 &  0.3   & 0.79       \\
log L$_{\rm Bol}^{\rm X}$ &  44.47  & 44.63  &  0.3   & 0.79            \\
log L$_{\rm Bol}^{\rm [OIII]}$ & 44.64   & 45.01 & 0.2 & 0.99         \\
log M$_{\rm BH}$ &  7.18  & 7.29 & 0.5 & 0.16            \\
log ${\lambda}_{\rm X}$ &  -2.06  & -1.70 & 0.2 & 0.98            \\
log ${\lambda}_{\rm [OIII]}$ &  -1.76  & -1.38 & 0.3 & 0.75         \\    \hline
\end{tabular} 
\\
Kolmogorov - Smirnov two sample test examines the hypothesis that two samples comes from same distribution.
D = Sup x $|$S1(x) - S2(x)$|$ is the maximum difference between the cumulative distributions of two samples S1(x) and S2(x), respectively.
\end{table}
\begin{figure*}
\centering
\includegraphics[angle=0,width=6.8cm,height=5.6cm]{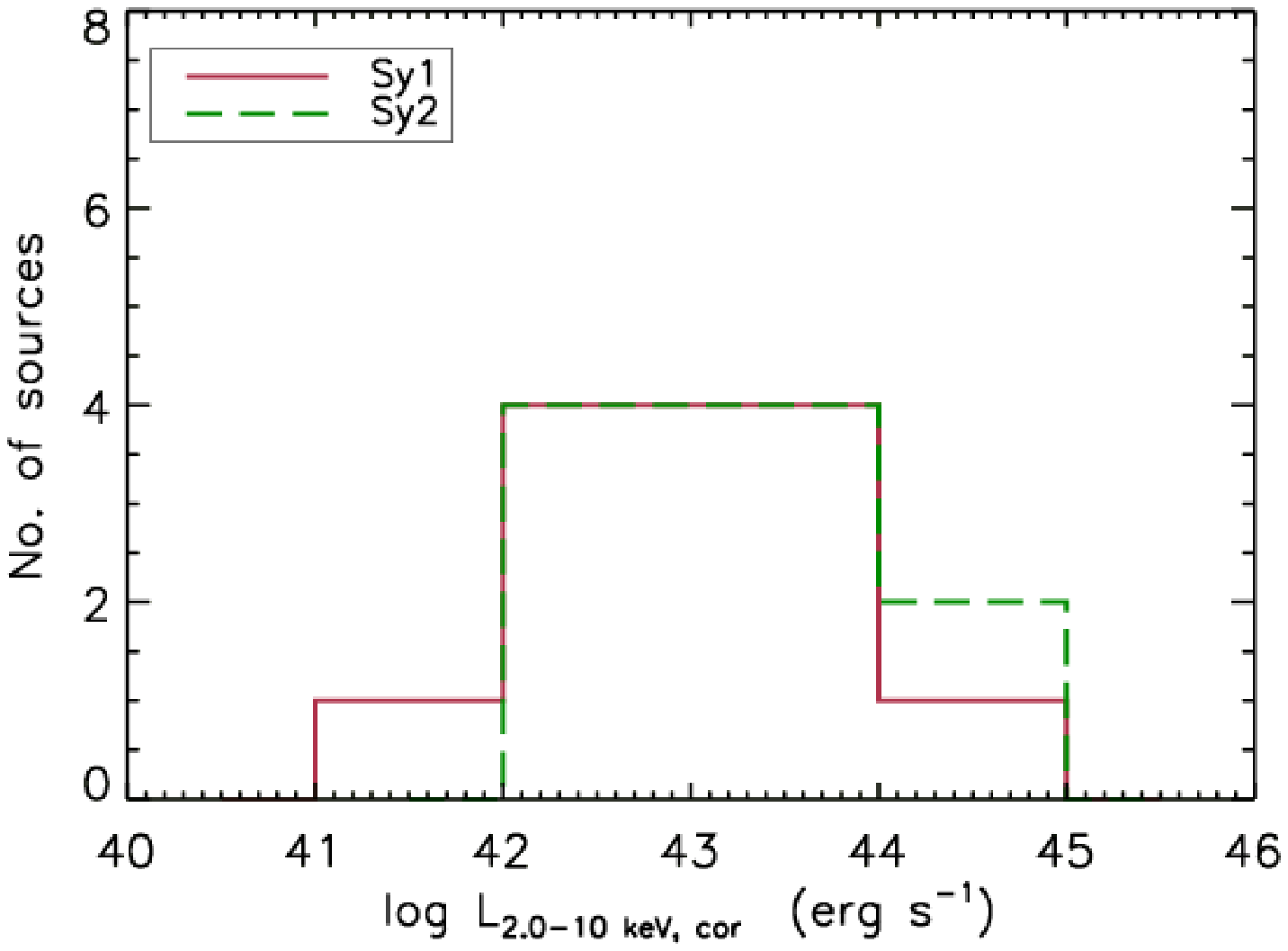}{\includegraphics[angle=0,width=6.8cm,height=5.6cm]{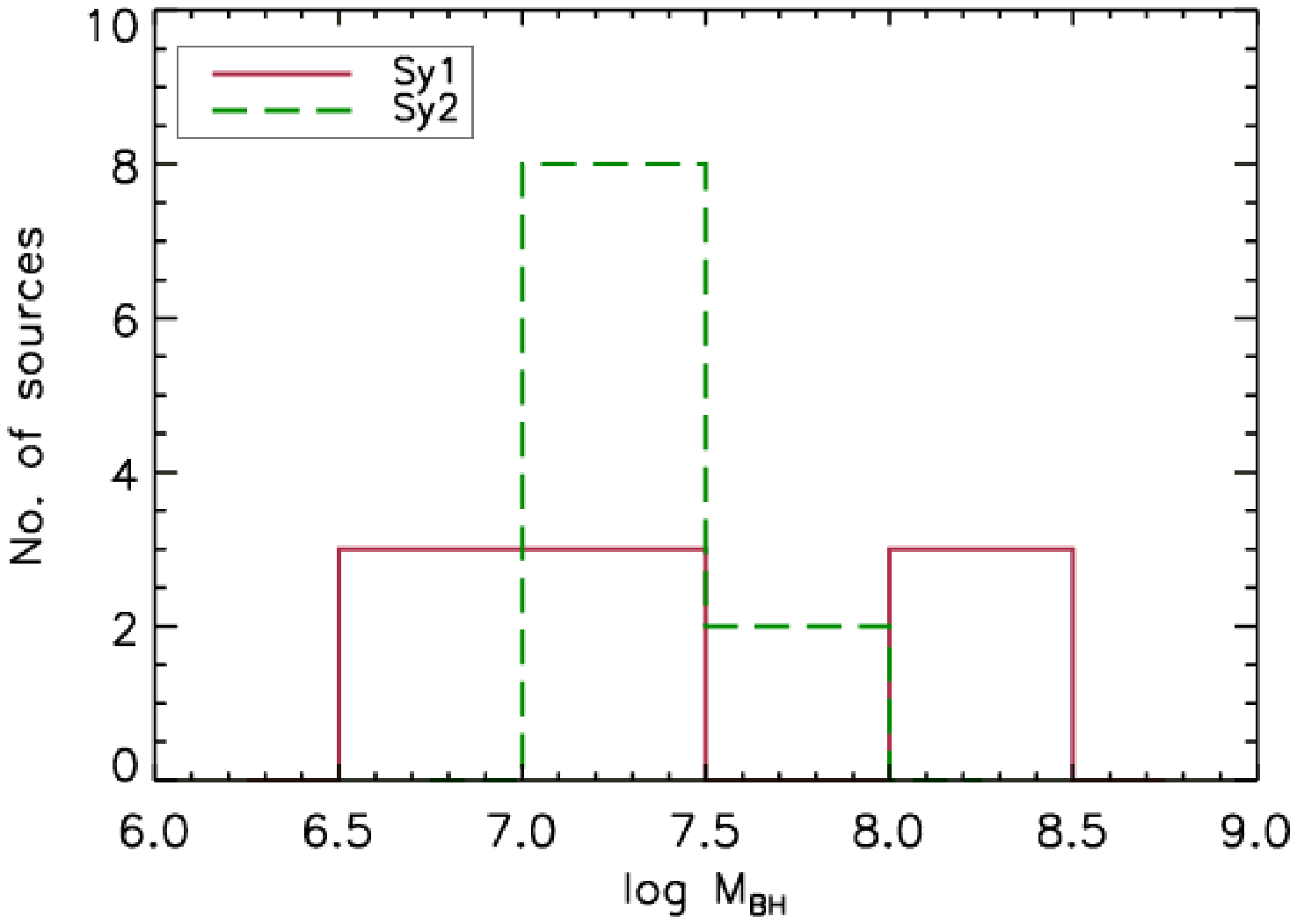}}
\caption{{\it Left}: Distributions of the absorption-corrected 2.0 - 10 keV X-ray luminosities (L$_{\rm 2.0~-~10~keV}$) for the two Seyfert subtypes. {\it Right}: Distribution of the black hole masses (in M$_{\odot}$ units) for the two Seyfert subtypes.}
\end{figure*}
\begin{figure*}
\centering
\includegraphics[angle=0,width=6.8cm,height=5.6cm]{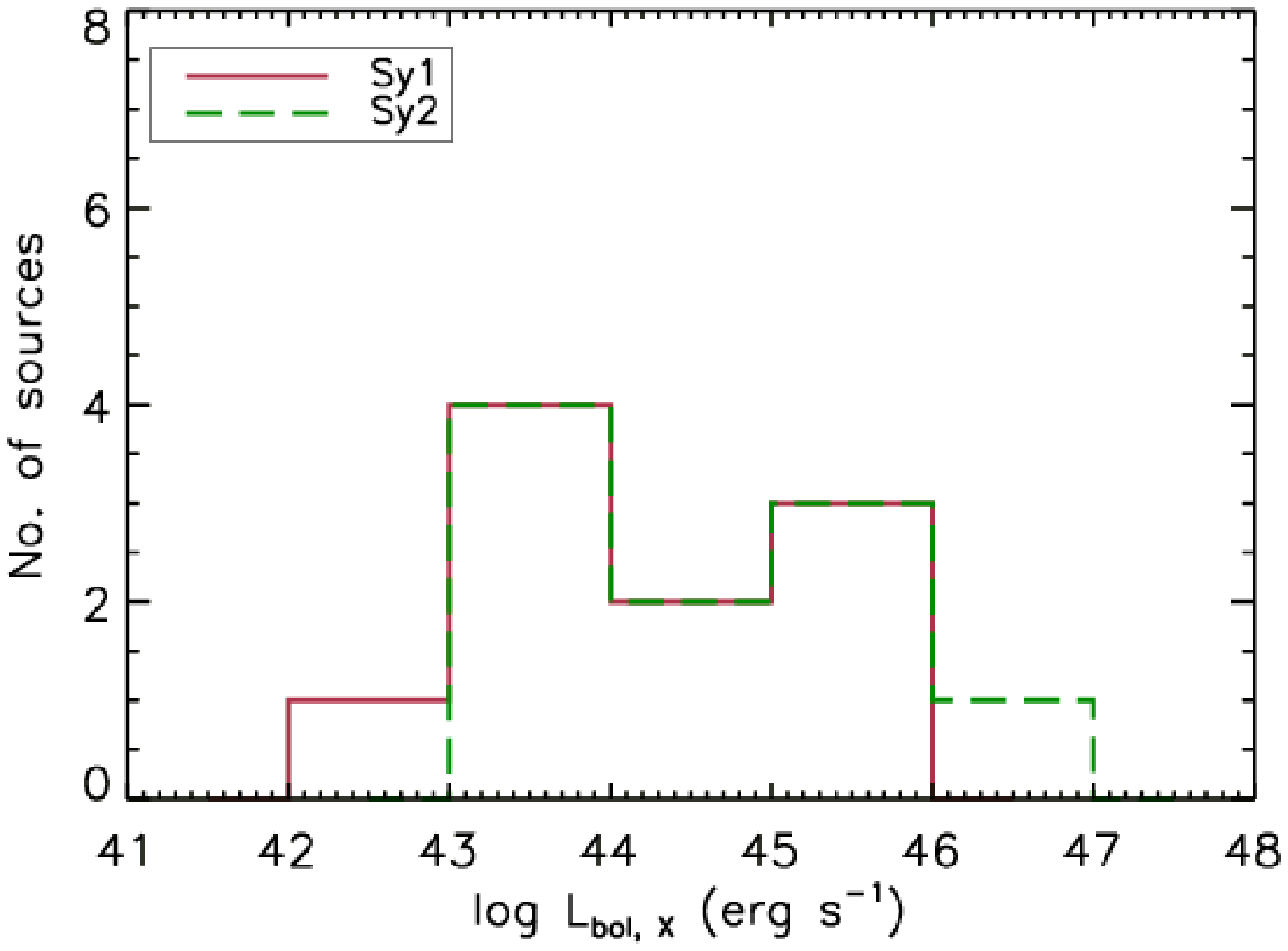}{\includegraphics[angle=0,width=6.8cm,height=5.6cm]{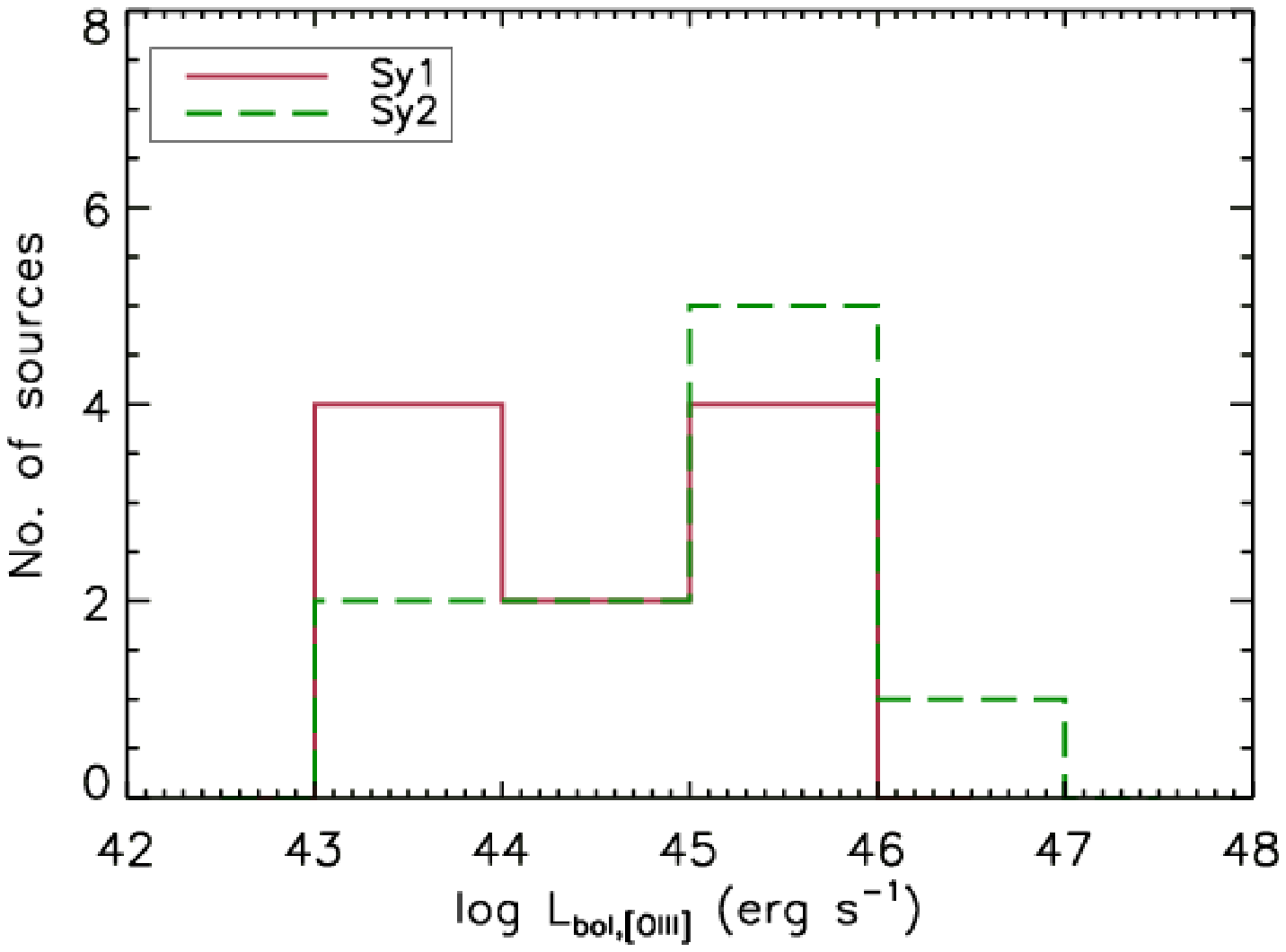}}
\caption{Distributions of the bolometric luminosities (L$_{\rm Bol}$) for the two Seyfert subtypes. {\it Left}: L$_{\rm Bol}$ is derived from 
L$_{\rm 2.0~-~10~keV}^{\rm c}$, {\it Right}: L$_{\rm Bol}$ is derived from L$_{\rm [OIII]}$.}
\end{figure*}
\begin{figure}
\centering
\includegraphics[angle=0,width=7.0cm,height=6.0cm]{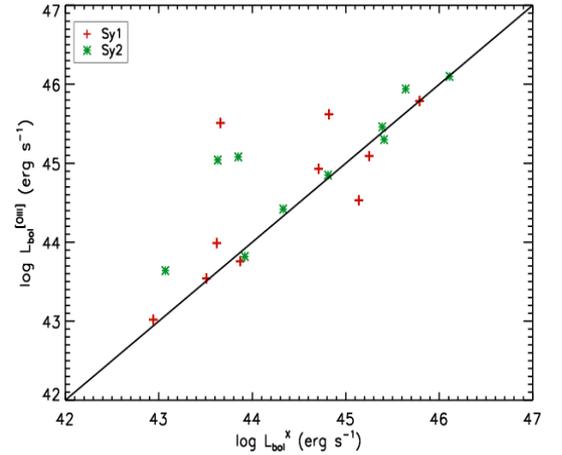}
\caption{Comparison of the bolometric luminosities (L$_{\rm Bol}$) obtained from L$_{\rm 2.0~-~10~keV}$ and L$_{\rm [OIII]}$ respectively. 
The solid line represents L$_{\rm Bol}^{\rm X}$ = L$_{\rm Bol}^{\rm [OIII]}$ line.}
\end{figure}
\begin{figure*}
\centering
\includegraphics[angle=0,width=6.8cm,height=5.6cm]{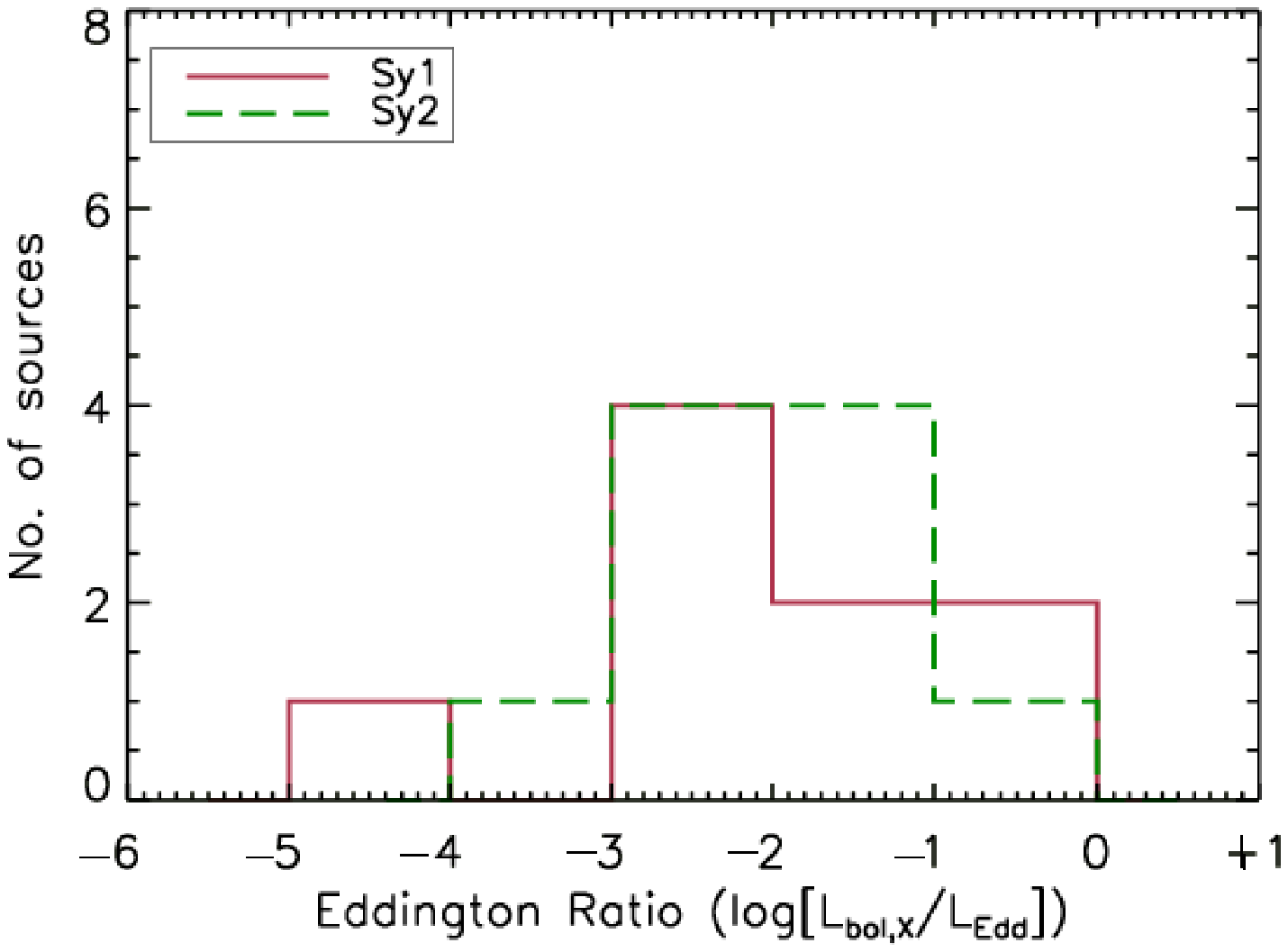}{\includegraphics[angle=0,width=6.8cm,height=5.6cm]{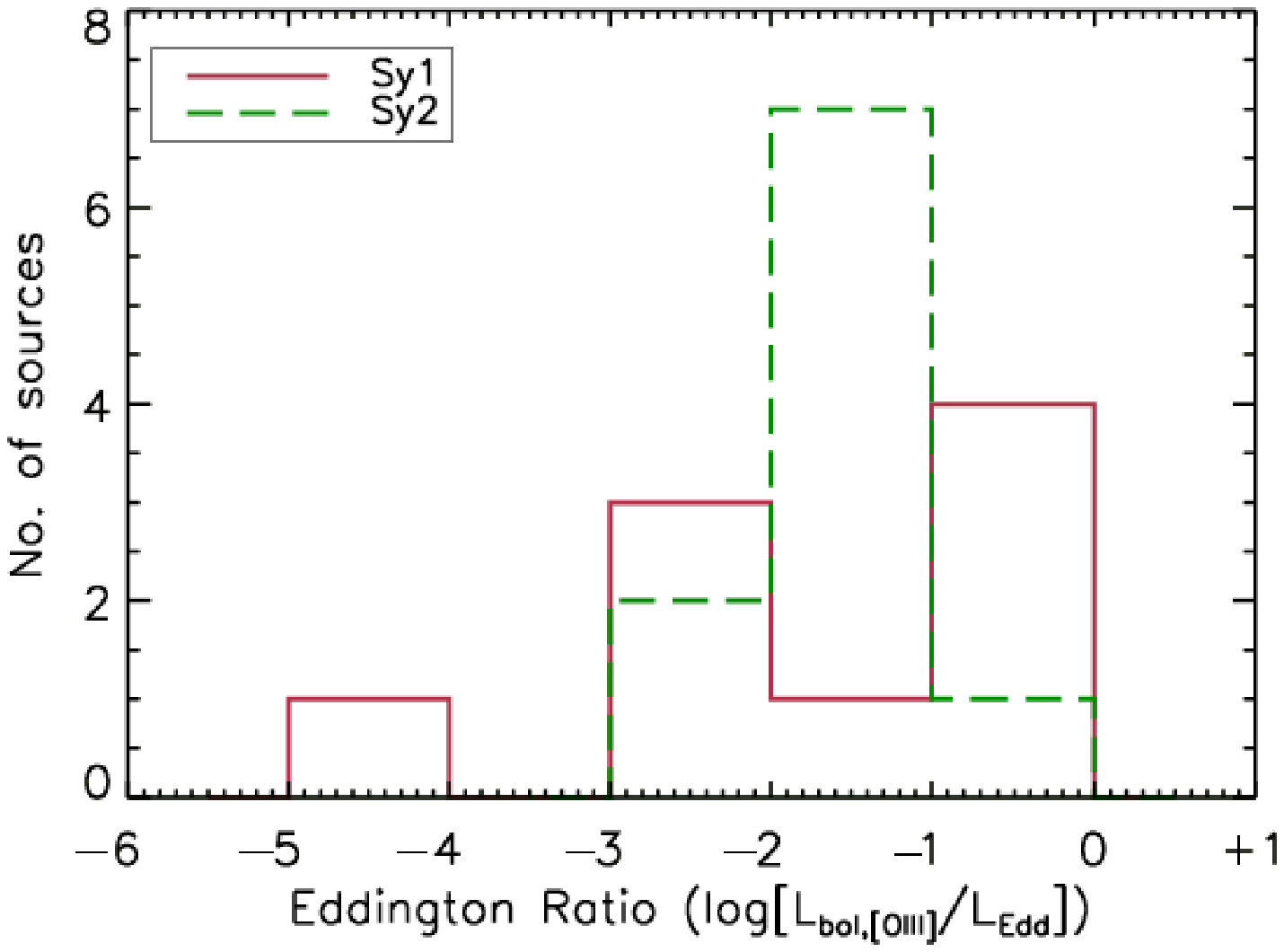}}
\caption{Distributions of the Eddington ratios for the two Seyfert subtypes. {\it Left}: L$_{\rm Bol}$ is derived from L$_{\rm 2.0~-~10~keV}$, 
{\it Right}: L$_{\rm Bol}$ is derived from L$_{\rm [OIII]}$.}
\end{figure*}
\section{Discussion}
Table 1 lists the absorption-corrected 2.0 - 10 keV X-ray luminosities (L$_{\rm 2.0~-10~keV}^{\rm c}$), the extinction-corrected 
[OIII] 5007$\mbox {\AA}$ line luminosities (L$_{\rm [OIII]}^{\rm c}$), the bolometric luminosities derived from the X-ray as well as the [OIII] luminosities (L$_{\rm Bol}^{\rm x}$, L$_{\rm Bol}^{\rm [OIII]}$), the black hole masses (M$_{\rm BH}$) and the Eddington ratios (${\lambda}_{\rm X}$, ${\lambda}_{\rm [OIII]}$) 
for the Seyfert type 1s and type 2s of our sample.
In following subsections we discuss the statistical comparisons of the nuclear properties of the two Seyfert subtypes with the primary aim to 
examine the validity of Seyfert unification scheme. 
\subsection{Comparison of absorption-corrected 2.0 - 10 keV X-ray luminosities}
In the unification scenario, the X-ray emitting AGN in Seyfert type 2s is viewed through the obscuring torus and therefore 
Seyfert type 2s are expected to show lower observed 2.0 - 10 keV X-ray luminosities than type 1s 
as long as the obscuring column density is sufficiently high (N$_{\rm H}$ $>$ 10$^{22}$ cm$^{-2}$). 
We have shown in paper-I that the Seyfert type 2s compared to type 1s, show systematically lower observed soft (0.5 - 2.0 keV) 
and hard (2.0 - 10 keV) X-ray luminosities, consistent with the unification scheme. 
Since the differences between the two Seyfert subtypes are due to the differing orientations of the obscuring torus and therefore if 2.0 - 10 keV X-ray luminosities are corrected for the line-of-sight absorption, 
both Seyfert type 1s and type 2s are expected to show similar 2.0 - 10 keV luminosity distributions. 
Figure 1 shows the 2.0 - 10 keV absorption-corrected X-ray luminosity distributions for the two Seyfert subtypes of our sample.
In our sample, the type 1s have L$_{\rm 2.0~-10~keV}^{\rm c}$ in the range of $\sim$ 8.3 $\times$ 10$^{41}$ to $\sim$ 1.6 $\times$ 10$^{44}$ 
erg s$^{-1}$ with the median value $\sim$ 1.5 $\times$ 10$^{43}$ erg s$^{-1}$, while for type 2s, L$_{\rm 2.0~-10~keV}^{\rm c}$ ranges from 
$\sim$ 1.1 $\times$ 10$^{42}$ to $\sim$ 2.8 $\times$ 10$^{44}$ erg s$^{-1}$ with the median value $\sim$ 2.0 $\times$ 10$^{43}$ erg s$^{-1}$.
The two sample KS test confirms that the Seyfert type 1s and type 2s have similar 2.0 - 10 keV absorption-corrected X-ray luminosity 
distributions, consistent with the unification scheme. 
The KS test shows that there is $\sim$ 79$\%$ probability that the L$_{\rm 2.0~-10~keV}^{\rm c}$ distributions of type 1 and type 2 Seyferts 
are drawn from the same parent population (Table 2). 
The absorption-corrected 2.0 - 10 keV X-ray luminosities for our sample Seyferts are similar to the ones reported in the previous studies 
\citep{Cappi06}.
The absorption-corrected 2.0 - 10 keV X-ray luminosity can be considered as the representative of AGN power as long as the X-ray photons 
in the 2.0 - 10 keV band are not substantially diminished by other than photoelectric absorption and 
are not contaminated by non-AGN emission.
Circumnuclear starburst is reported to be present in some of our sample sources ({\eg}MRK~477, \citep{Heckman97}; NGC~5135, \citep{Gonzalez98}), 
however, it has been argued that the circumnuclear starburst mainly contribute to the soft X-ray 0.5 - 2.0 keV band and 
its contamination to 2.0 - 10 keV keV energy band is not substantial \citep{Levenson04}.
\subsection{Comparison of AGN bolometric luminosities}
The bolometric luminosity of AGN represents the rate of energy emitted over the entire electromagnetic waveband 
by the accreting black hole. 
AGN bolometric (L$_{\rm Bol}$) depends on the mass accretion rate and the radiative efficiency of the accreting matter.
Therefore, if Seyfert type 1s and type 2s are 
intrinsically similar as hypothesized by the unification scheme, then both the subtypes should show similar distributions of the bolometric luminosity.
As discussed in the section 3, we have estimated the bolometric luminosities of our sample sources using the absorption-corrected 
2.0 - 10 keV X-ray luminosity as well as the [OIII] line luminosity. 
Figure 2 shows the distributions of bolometric luminosities (L$_{\rm Bol}^{\rm X}$ and L$_{\rm Bol}^{\rm [OIII]}$) for the Seyfert type 1s and type 2s.
In our sample, Seyfert type 1s have L$_{\rm Bol}^{\rm X}$ ranging from $\sim$ 8.7 $\times$ 10$^{42}$ to $\sim$ 6.2 $\times$ 10$^{46}$ erg s$^{-1}$ with 
the median value $\sim$ 3.0 $\times$ 10$^{44}$ erg s$^{-1}$, while type 2s have L$_{\rm Bol}^{\rm X}$ in the range of 
$\sim$ 1.2 $\times$ 10$^{43}$ to $\sim$ 1.3 $\times$ 10$^{46}$ erg s$^{-1}$ with 
the median value $\sim$ 4.3 $\times$ 10$^{44}$ erg s$^{-1}$. 
The bolometric luminosities derived from the [OIII] luminosities (L$_{\rm Bol}^{\rm [OIII]}$) also span over similar range.
The L$_{\rm Bol}^{\rm [OIII]}$ for type 1s spans over $\sim$ 1.0 $\times$ 10$^{43}$ to $\sim$ 6.2 $\times$ 10$^{46}$ erg s$^{-1}$ with 
the median value $\sim$ 4.4 $\times$ 10$^{44}$ erg s$^{-1}$, while for type 2s, it ranges from $\sim$ 4.4 $\times$ 10$^{44}$ to 
$\sim$ 1.3 $\times$ 10$^{46}$ erg s$^{-1}$ with the median value $\sim$ 1.0 $\times$ 10$^{45}$ erg s$^{-1}$. 
The two sample KS test shows that the two Seyfert subtypes of our sample have similar distributions of bolometric luminosities 
(L$_{\rm Bol}^{\rm X}$ and L$_{\rm Bol}^{\rm [OIII]}$). There is $\sim$ 79$\%$ and $\sim$ 99$\%$ probability, respectively that 
the L$_{\rm Bol}^{\rm X}$ and L$_{\rm Bol}^{\rm [OIII]}$ distributions of Seyfert type 1s and type 2s are drawn from the same parent population 
(Table 2).
The similar distributions of the bolometric luminosities for the Seyfert type 1s and type 2s of our sample 
are in the lines as expected from the unification scheme. 
The bolometric luminosities (derived from the X-ray as well as the [OIII] luminosities) for the Seyferts of our sample are in agreement with the ones 
reported in the previous studies \citep{Ho09}.
\\
\subsection{Comparison of black hole masses}
Black hole mass is one of the fundamental parameters of AGN and if Seyfert type 1s and type 2s differ only due to the orientation of obscuring torus 
and host similar AGN, then, they are expected to have similar distributions of the supermassive black hole masses.
We note that in our sample, black hole masses (M$_{\rm BH}$) for Seyfert type 1s span over 
$\sim$ 3.2 $\times$ 10$^{6}$ to $\sim$ 1.2 $\times$ 10$^{8}$ M$_{\odot}$ with the median value $\sim$ 1.5 $\times$ 10$^{7}$ M$_{\odot}$, while 
for type 2s, it ranges from $\sim$ 1.4 $\times$ 10$^{7}$ to $\sim$ 7.4 $\times$ 10$^{7}$ M$_{\odot}$ with 
the median value $\sim$ 1.9 $\times$ 10$^{7}$ M$_{\odot}$.
The comparison shows that the distributions of the black hole masses for the type 1s and type 2s of our sample are not 
significantly different (Figure 1), which is consistent with the unification scheme. 
The low p-value $\sim$ 0.16 (Table 2) in the two sample KS test could be attributed to the small sample size. 
Our results are broadly in agreement with the previous studies \citep{Panessa06,Bian07,Middleton08} which report that the black hole mass 
(M$_{\rm BH}$) distributions are similar for Seyfert type 1s and type 2s.
\subsection{Comparison of Eddington ratios}
Eddington ratio ($\lambda$ = L$_{\rm Bol}$/L$_{\rm Edd}$) {\ie}AGN bolometric luminosity normalized with Eddington luminosity 
can be used to characterize the level of nuclear activity \citep{Ho08,Ho09}. 
For our sample sources we have two estimates of Eddington ratios {\ie}${\lambda}_{\rm X}$, where absorption-corrected 2.0 - 10 keV X-ray 
luminosity is used to estimate the AGN bolometric luminosity (L$_{\rm Bol}$) and ${\lambda}_{\rm [OIII]}$, where [OIII] 
luminosity is used to estimate L$_{\rm Bol}$.
Figure 4 shows the distributions of ${\lambda}_{\rm X}$ and ${\lambda}_{\rm [OIII]}$ for the two Seyfert subtypes. 
We note that for the Seyfert type 1s, ${\lambda}_{\rm X}$ ranges from $\sim$ 6.3 $\times$ 10$^{-5}$ to $\sim$ 3.5 $\times$ 10$^{-1}$ 
with the median value $\sim$ 8.7 $\times$ 10$^{-3}$, while for type 2s, it ranges 
from $\sim$ 5.1 $\times$ 10$^{-4}$ to $\sim$ 6.2 $\times$ 10$^{-1}$ with the median value $\sim$ 2.0 $\times$ 10$^{-2}$.
The two sample KS test shows that there is $\sim$ 98$\%$ probability that the ${\lambda}_{\rm X}$ distributions of Seyfert type 1s and type 2s 
are drawn from the same parent population (Table 2). 
The distributions of ${\lambda}_{\rm [OIII]}$ are also similar (p-value $\sim$ 0.75) for the two Seyfert subtypes.
For Seyfert type 1s, ${\lambda}_{\rm [OIII]}$ spans over $\sim$ 7.6 $\times$ 10$^{-5}$ to $\sim$ 4.6 $\times$ 10$^{-1}$ 
with the median value $\sim$ 1.7 $\times$ 10$^{-3}$, while for type 2s, it spans over 
$\sim$ 1.9 $\times$ 10$^{-3}$ to $\sim$ 6.2 $\times$ 10$^{-1}$ with the median value $\sim$ 4.2 $\times$ 10$^{-2}$.
The similar distributions of Eddington ratios for the two Seyfert subtypes implies that 
type 1s and type 2s have similar level of accretion and therefore, both the subtypes are intrinsically similar.
Some of the literature studies have reported that the Seyfert type 2 galaxies accrete 
at lower Eddington ratios than type 1 Seyfert galaxies although black hole mass distributions for the two subtypes are similar \citep{Panessa06,Middleton08}. Our results are in contrary to these studies which report that the Seyfert type 2s are accreting at lower Eddington ratios in compared to type 1s. 
We note that while discussing the difference in the Eddington ratio distributions of the two Seyfert subtypes, \cite{Panessa06} 
mention the points of caveats, {\eg}they used a constant correction factor to estimate the bolometric luminosity 
which may not be appropriate with the fact that 
the bolometric luminosity depends on the shape of the spectral energy distribution which could differ from high to low luminosity 
AGN \citep{Ho99,Marconi04}. Also, the absorption-corrected X-ray luminosity distributions for type 1s and 2s of their sample are substantially different 
which may consequently affect the distributions of Eddington ratios. 
\cite{Middleton08} sample is based on hard X-ray observations {\ie}{\em INTEGRAL}-IBIS, {\em BeppoSAX}-PDS, {\em CGRO}-OSSE, and therefore 
it may be biased to towards relatively bright sources. \cite{Middleton08} emphasized that Seyfert type 1s and type 2s need to be matched in 
intrinsic properties such as Eddington ratio in order to explore the orientation based differences. 
In fact, we follow the similar approach by using a sample in which two Seyfert subtypes are matched in orientation-independent 
properties \citep{Lal11}.
\par
Seyfert galaxies of our sample have Eddington ratios much lower compared to luminous AGNs 
({\ie}quasars showing Eddington ratios $>$ 10$^{-1}$ 
\citep{Shemmer06}) suggesting that the low luminosity of Seyfert AGN is likely due to low accretion rate.
The AGN bolometric luminosities of the Seyfert galaxies (considering both the subtypes together) 
of our sample range from $\sim$ 8.7 $\times$ 10$^{42}$ erg s$^{-1}$ to $\sim$ 6.2 $\times$ 10$^{46}$ erg s$^{-1}$ 
with the median value of $\sim$ 3.6 $\times$ 10$^{44}$ erg s$^{-1}$. If this AGN emission is produced by a canonical optically thick, geometrically 
thin accretion disk \citep{Shakura73} that radiates at 
L$_{\rm acc}$ = ${\eta}{\dot{\rm M}}{\rm c^{2}}$ = 5.7 $\times$ 10$^{45}$ ($\eta$/0.1) (${\dot{\rm M}}$/M$_{\odot}$ year$^{-1}$) erg s$^{-1}$ 
(see, \cite{Ho09}), then we expect a typical mass accretion rate in the range of 
${\dot{\rm M}}$ $\sim$ 1.5 $\times$ 10$^{-3}$ - 11 M$_{\odot}$ year$^{-1}$ with the 
median value of 6.3 $\times$ 10$^{-2}$ M$_{\odot}$~year$^{-1}$, assuming radiative efficiency $\eta$ = 0.1.
\cite{Ho09} argued that the fuel required for the accretion rate ${\dot{\rm M}}$ $\sim$ 10$^{-3}$ M$_{\odot}$ yr$^{-1}$ 
can readily be supplied by the mass loss from evolved stars present in the bulge and the diffuse hot gas present in the circumnuclear region. 
The additional supply of fuel such as higher mass loss rates from the circumnuclear starburst, dissipation from larger scales nuclear 
bar or spirals and stellar tidal disruption is likely to present and therefore higher accretion rates are expected than that shown by 
low luminosity Seyferts. Radiatively inefficient accretion flows (RIAFs; \citep{Narayan98,Quataert01}) are generally invoked to explain 
the observed low Eddington ratios in low luminosity AGN.
The relation between accretion rate and luminosity in RIAFs can be expressed as 
${\dot{\rm m}}$ $\approx$ 0.7 ($\alpha$/0.3)(L$_{\rm bol}$/L$_{\rm Edd}$)$^{1/2}$, where 
${\dot{\rm m}}$ = ${\dot{\rm M}}$/${\dot{\rm M}_{\rm Edd}}$ and ${\alpha}$ is viscosity parameter assumed to be $\approx$ 0.1 - 0.3 (see, \cite{Ho09}). 
The Seyfert galaxies of our sample have L$_{\rm bol}$/L$_{\rm Edd}$ $\sim$ 6.3 $\times$ 10$^{-5}$ - 6.2 $\times$ 10$^{-1}$ and this implies 
${\dot{\rm m}}$ $\approx$ 1.9 $\times$ 10$^{-3}$ - 1.8 $\times$ 10$^{-2}$. These values of accretion rates lie 
well within the regime of optically thin RIAFs (${\dot{\rm m}}$ $\leq$ ${\dot{\rm m}}_{\rm crit}$ $\approx$ ${\alpha}^{2}$ $\approx$ 0.1 \citep{Narayan98}.
\section{Conclusions}
\begin{itemize}
\item We have argued for the importance of sample selection in testing the predictions of the Seyfert unification scheme and 
showed that the distributions of the nuclear properties {\ie}the absorption-corrected 2.0 - 10 keV X-ray luminosities, 
the bolometric luminosities, the black hole masses and the Eddington ratios are similar for the two Seyfert subtypes of 
a sample based on orientation-independent 
properties. Our results on the statistical comparison of the nuclear properties of the Seyfert type 1s and type 2s are 
consistent with the orientation and obscuration based Seyfert unification scheme.   
\item The absorption-corrected 2.0 -10 keV X-ray luminosities are in the range of 
$\sim $10$^{42}$ - 10$^{44}$ erg s$^{-1}$ for both the Seyfert subtypes, with similar median values. 
The estimated bolometric luminosities derived from the X-ray as well as the [OIII] luminosities are in range of 
$\sim $10$^{43}$ - 10$^{46}$ erg s$^{-1}$ for both the Seyfert subtypes, with similar median values.
\item Both the Seyfert subtypes accrete with sub-Eddington rates {\ie}Eddington ratios range $\sim$ 10$^{-4}$ - 10$^{-1}$ 
in compared to luminous AGNs ({\eg}quasars with Eddington ratio $\sim$ 1). 
The plausible explanation for the substantially low Eddington ratios may be the radiatively  inefficient accretion such as 
advection dominated accretion flow and their variants \citep{Narayan94}.
\end{itemize}
\bibliographystyle{aa}
\bibliography{VeereshSinghXMMpaper}
\end{document}